\documentclass[square,comma,sort&compress]{elsart}
\usepackage{amsfonts}
\usepackage{amsmath}
\usepackage{amssymb}
\usepackage{natbib}
\usepackage{epsfig}
\usepackage{rotating}
\usepackage{graphicx}

\newcommand{\QUESTION}[1]{}
\newcommand{\ANSWER}[1]{}

\begin{document}

\runauthor{Ellero}
\begin{frontmatter}
\title{Unique representation of an inverse-kinetic theory for incompressible Newtonian fluids}
\author[UTS]{M. Tessarotto}
\author[SYD]{and M. Ellero}

\address[UTS]{Department of Mathematics
and Informatics, University of Trieste, Trieste, Italy and
Consortium for Magnetofluid Dynamics, Trieste, Italy}

\address[SYD]{School of Aerospace, Mechanical and Mechatronic Engineering, The
University of Sydney, NSW 2006, Australia}

%
\thanks[maxtex]{Corresponding author: email: marco.ellero@cmfd.univ.trieste.it}

\begin{abstract}
Fundamental aspects of inverse kinetic theories for the incompressible
Navier-Stokes equations [Ellero and Tessarotto, 2004, 2005]
include the possibility of defining uniquely the kinetic equation
underlying such models and furthermore, the construction of a
kinetic theory implying also the energy equation. The latter
condition is consistent with the requirement that fluid fields
result classical solutions of the fluid equations. These issues
appear of potential relevance both from the mathematical viewpoint
and for the physical interpretation of the theory. Purpose of this
work is to prove that under suitable prescriptions the inverse
kinetic theory can be determined to satisfy such requirements.
\newline

PACS: 47.27.Ak, 47.27.Eq, 47.27.Jv

\end{abstract}

\begin{keyword}%

Incompressible Navier-Stokes equations: kinetic theory;
Fokker-Planck equation.
\end{keyword}
\end{frontmatter}

\section{Introduction}

An aspect of fluid dynamics is represented by the class of so-called \textit{%
inverse problems}, involving the search of model kinetic theories able to
yield a prescribed complete set of fluid equations advancing in time a
suitable set of fluid fields. This is achieved by means of suitable
velocity-moments of an appropriate kinetic distribution function $f(\mathbf{%
r,v,}t)$. Among such model theories, special relevance pertains to those in
which the state of isothermal incompressible fluids is described
self-consistently by suitable fluid fields to be considered as classical
(i.e., strong) solutions of the corresponding fluid equations (\emph{%
regularity assumption}). In this case the relevant fluid fields are the mass
density, fluid velocity and fluid pressure $\left\{ \rho ,\mathbf{V,}%
p\right\} $ which are required to be classical solutions of the so-called%
\emph{\ incompressible Navier-Stokes equations} (INSE)

\begin{eqnarray}
\frac{\partial }{\partial t}\rho +\nabla \cdot \left( \rho \mathbf{V}\right)
&=&0,  \label{INSE-1} \\
\text{ \ \ \ \ \ \ \ \ \ \ \ \ \ \ \ \ \ \ \ \ \ \ \ \ }\rho \frac{D}{Dt}%
\mathbf{V}+\mathbf{\nabla }p+\mathbf{f}-\mu \nabla ^{2}\mathbf{V} &=&\mathbf{%
0},  \label{INSE-2} \\
\nabla \cdot \mathbf{V} &=&0,  \label{INSE-3} \\
\rho (\mathbf{r,}t) &>&0,  \label{INSE-4} \\
p(\mathbf{r,}t) &\geq &0,  \label{INSE-5} \\
\rho (\mathbf{r,}t\mathbb{)} &=&\rho _{o}>0.  \label{INSE-6b}
\end{eqnarray}%
The first three equations (\ref{INSE-1}),(\ref{INSE-2}) and (\ref{INSE-3}),
denoting respectively the continuity, Navier-Stokes and isochoricity
equations, are assumed to be satisfied in the open set $\Omega \subseteq
\mathbb{R}^{3}$ (fluid domain) and in a possibly bounded time interval $%
I\subset \mathbb{R},$ while the last three inequalities, (\ref{INSE-4})-(\ref%
{INSE-6b}) apply also in the closure of the fluid domain $\overline{\Omega }%
\equiv \Omega \cup \delta \Omega .$ Here the notation is standard. Hence $%
\frac{D}{Dt}=\frac{\partial }{\partial t}+\mathbf{V\cdot \nabla }$ is the
convective derivative, $\mathbf{f(r,}t)$ denotes the volume force density
acting on the fluid element and $\mu \equiv \nu \rho _{o}>0$ is the constant
fluid viscosity, $\nu =\mu /\rho _{o}$ being the related kinematic
viscosity. It is assumed that the fluid fields and $\mathbf{f(r,}t)$ are
suitably smooth to belong to the functional setting

\begin{equation}
\left\{
\begin{array}{l}
\mathbf{V}(\mathbf{\mathbf{r,}}t),p(\mathbf{r,}t),\mathbf{\mathbf{f}}(%
\mathbf{\mathbf{r,v}},t\mathbb{)}\in C^{(0)}(\overline{\Omega }\times I), \\
\mathbf{V}(\mathbf{\mathbf{r,}}t),p(\mathbf{r,}t)\in C^{(2,1)}(\Omega \times
I), \\
\mathbf{f}(\mathbf{r,}t)\in C^{(1,0)}(\Omega \times I),%
\end{array}%
\right.  \label{NFS}
\end{equation}%
[where $C^{(i,j)}(\Omega \times I)\equiv C^{(i,)}(\Omega )\times C^{(j)}I),$
with $i,j\in \mathbb{N}$]. \ Hence, $\left\{ \rho ,\mathbf{V,}p\right\} $
are classical solutions of INSE so that the energy equation, obtained by
taking the scalar product of the Navier-Stokes equation by the fluid
velocity $\mathbf{V,}$ holds identically in the same domain%
\begin{equation}
\frac{D}{Dt}\frac{V^{2}}{2}+\frac{1}{\rho _{o}}\mathbf{V\cdot \nabla }p+%
\frac{1}{\rho _{o}}\mathbf{V\cdot f}-\nu \mathbf{V\cdot }\nabla ^{2}\mathbf{V%
}=0.  \label{energy equation}
\end{equation}%
The set of equations (\ref{INSE-1})-(\ref{INSE-6b}), together with (\ref%
{energy equation}), will be denoted in the sequel as \emph{extended INSE}. \

An inverse kinetic theory for INSE is therefore represented by a kinetic
equation of the form%
\begin{equation}
L(\mathbf{F})f=0,  \label{inverse kinetic equation}
\end{equation}%
$f(\mathbf{x,}t)$ being a kinetic distribution function, defined in the
extended phase space $\Gamma \times I,$ where $\Gamma =\Omega \times U$
(with $\Gamma $ and $U$ the phase and velocity spaces), $\overline{\Gamma }=%
\overline{\Omega }\times U$ its closure, while $\mathbf{x}\equiv (\mathbf{r,v%
})\in \Gamma $ and $\mathbf{v}\in U\equiv \mathbb{R}^{3}$ denote
respectively the state and velocity variables. The distribution $f(\mathbf{x,%
}t)$ is assumed suitably regular (i.e., suitably smooth in $\Gamma \times I$
and summable in $\Gamma )$ and to obey appropriate initial and boundary
conditions. 
$L(\mathbf{F})$ is an appropriate operator (to be denoted as \textit{Vlasov}
\textit{streaming operator}). This is defined in such a way that appropriate
velocity moments of (\ref{inverse kinetic equation}), are assumed to exist,
yield INSE. In particular, $L(\mathbf{F})$ can be defined in such a way to
allow that the inverse kinetic equation (\ref{inverse kinetic equation})
admits, as a particular solution, the local Maxwellian distribution%
\begin{equation}
f_{M}(\mathbf{x,}t;\mathbf{V,}p_{1})=\frac{\rho _{0}^{5/2}}{\left( 2\pi
\right) ^{\frac{3}{2}}p_{1}^{\frac{3}{2}}}\exp \left\{ -X^{2}\right\} .
\label{local Maxwellian distribution}
\end{equation}%
Here, the notation is standard \cite{Ellero2005}, thus
\begin{eqnarray}
\text{ \ \ \ \ \ \ \ \ \ \ \ \ \ \ \ \ \ \ \ \ \ }X^{2} &=&\frac{u^{2}}{%
v_{th}{}^{2}}, \\
v_{th}^{2} &=&2p_{1}/\rho _{o},
\end{eqnarray}%
$p_{1}$ being the kinetic pressure. Desirable features of the inverse
kinetic theory involve the requirement that, under suitable assumptions, the
functional form of the relevant inverse kinetic equation, yielding the INSE
equations, be uniquely defined. 
In addition, it might be convenient to impose also the validity of
additional fluid equations, such for example the energy equation (\ref%
{energy equation}). In fact, it is well known that the energy equation is
not satisfied by weak solutions of INSE and, as a consequence, also by
certain numerical schemes. 
Therefore, the validity of the inverse kinetic equation yields a necessary
condition for the existence of classical solutions for INSE.

Concerning the first point, the prescription of uniqueness on the kinetic
equation has to be intended in a suitably meaningful sense, i.e., to hold
under the requirement that the relevant set of fluid equations are fulfilled
identically by the fluid fields in the extended domain $\Omega \times I$.
This means that arbitrary contributions in the kinetic equation, which
vanish identically under a such an hypothesis, can be included in the same
kinetic equation. Consistently with the previous regularity assumption, here
we intend to consider, in particular, the requirement that the inverse
kinetic equation yields also the energy equation (\ref{energy equation}).

In a previous work \cite{Tessarotto2004,Ellero2005}, an explicit solution to
INSE has been discovered based on a continuous inverse kinetic theory,
adopting a Vlasov differential kinetic equation defined by a suitable
streaming operator $L.$ Basic feature of the kinetic equation is that,
besides yielding INSE as moment equations, it allows as particular solution
local kinetic Maxwellian equilibria for arbitrary fluid fluids $\left\{ \rho
_{o},\mathbf{V,}p\right\} $ which belong to the above functional setting (%
\ref{NFS})$.$ \ However, as pointed out in \cite{Ellero2005}, the inverse
kinetic equation defined in this way results parameter-dependent and hence
non-unique, even in the case of local Maxwellian kinetic equilibria. This
non-uniqueness feature may result as a potentially undesirable feature of
the mathematical model, since it prevents the possible physical
interpretation of the theory (in particular, of the mean-field force $%
\mathbf{F}$) and may result inconvenient from the numerical viewpoint since $%
\left\vert \alpha \right\vert $ may be chosen, for example, arbitrarily
large. \ Hence it would be highly desirable to eliminate it from the theory.

The purpose of this paper is twofold.

First we intend to prove that under suitable prescriptions the inverse
kinetic equation can be cast in an unique form, thus eliminating \ possible
parameter-dependences in the relevant streaming operator [$L(\mathbf{F})$]$.$%
This is achieved by analyzing the form of the streaming operator for
particular solutions (local Maxwellian kinetic equilibria). In this case the
the kinetic equation can be cast uniquely in an equivalent symmetrized form
represented by a Vlasov streaming operator.

As further development of the theory, it is shown that the streaming
operator can be suitably modified in such a way that the inverse kinetic
equation yields the extended INSE equations, i.e., besides the
incompressible Navier-Stokes equations also the energy equation. In
particular we intend to prove that the mean-field force $\mathbf{F}$ can be
uniquely defined in such a way that both kinetic equilibrium and moment
equations yield uniquely such equations.

The scheme of the presentation is as follows. In Sec.2 the inverse kinetic
equation developed in \cite{Ellero2005} is recalled and the non-uniqueness
feature of the mean-field force $\mathbf{F}$ is analyzed. In Sec.3 an
equivalent representation of the streaming operator is introduced which
permits to define uniquely $\mathbf{F}$. As a result, a uniqueness theorem
is obtained for the streaming operator $L(\mathbf{F})$. Finally in Sec.4 an
extension of the inverse kinetic theory is presented which provides a
solution also for the energy equation, besides the incompressible
Navier-Stokes equations. The formulation of the inverse kinetic equation for
the extended set of fluid equations is obtained by a suitable redefinition
of the mean-field force $\mathbf{F}$. Also in such a case the vector field $%
\mathbf{F}$ is proven to be unique.

\section{Non-uniqueness of the streaming operator}

Goal of this Section is to investigate the kinetic equation developed in {%
\cite{Ellero2005}} to analyze its non-uniqueness features. We start
recalling the inverse kinetic equation, which is assumed to be of the form (%
\ref{inverse kinetic equation}) or $L(\mathbf{F})\widehat{f}=0,$ $\widehat{f}%
(\mathbf{x,}t)$ being the normalized kinetic distribution function
associated to the kinetic distribution function $f(\mathbf{x,}t),$
\begin{equation}
\widehat{f}(\mathbf{x,}t)\equiv f(\mathbf{x,}t)/\rho _{o}.
\label{probability density}
\end{equation}%
In particular, the streaming operator $L$ is assumed to be realized by a
differential operator of the form
\begin{eqnarray}
&&\left. L(\mathbf{F})=\frac{\partial }{\partial t}+\mathbf{v\cdot }\frac{%
\partial }{\partial \mathbf{r}}+\right.  \label{streaming operator} \\
&&\left. +\frac{\partial }{\partial \mathbf{v}}\cdot \left\{ \mathbf{F}%
\right\} \right.
\end{eqnarray}%
and $\mathbf{F(r,v,}t;f)$ an appropriate vector field (\textit{mean-field
force}) defined by Eq.(\ref{F non-Maxwellian}) (see Appendix A) in terms of
vector fields $\mathbf{F}_{0}$ and $\mathbf{F}_{1}$. As a consequence, both $%
\mathbf{F}_{0}$ and $\mathbf{F}_{1}$are functionally dependent on the form
of the kinetic distribution function $f(\mathbf{x,}t)$. In particular,
requiring that $\mathbf{F}$ depends on the minimal number of velocity
moments 
(see below), it is defined by Eqs. (\ref{F non-Maxwellian}),(\ref{F0
non-maxwellian case}) and (\ref{F1 non-Maxwellian case}), given in Appendix
A. \ Supplemented with suitable initial and boundary conditions
and subject to suitable smoothness assumptions for the kinetic distribution
function $f(\mathbf{x,}t)$, several important consequences follow \cite%
{Ellero2005}:

\begin{itemize}
\item the fluid fields $\left\{ \rho _{o},\mathbf{V,}p\right\} $ can be
identified in the whole fluid domain $\Omega $ with suitable velocity
moments (which are assumed to exist) of the kinetic distribution function $f(%
\mathbf{x,}t)$ [or equivalent $\widehat{f}(\mathbf{x,}t)$], of the form%
\begin{equation}
M_{G}(r,t)=\int d^{3}vG(\mathbf{x},t)f(\mathbf{x,}t),
\label{velocity moments}
\end{equation}%
where \ $G(\mathbf{x},t)=1,\mathbf{v,}E\equiv \frac{1}{3}u^{2},\mathbf{v}E,$
$\mathbf{uu,}$ and%
\begin{equation}
\mathbf{u}\mathbb{\equiv }\mathbf{v}-\mathbf{V}(\mathbf{r,}t)
\end{equation}%
is the relative velocity.\textbf{\ }Thus, we require respectively
\begin{equation}
\rho _{o}=\int d^{3}vf(\mathbf{x,}t),  \label{moment-1}
\end{equation}%
\begin{equation}
\mathbf{V}(\mathbf{r,}t)=\frac{1}{\rho }\int d^{3}v\mathbf{v}f(\mathbf{x,}t),
\label{moment-2}
\end{equation}%
\begin{equation}
p\mathbf{(r,}t)=p_{1}\mathbf{(r,}t)-P_{o},  \label{pressure}
\end{equation}%
$p_{1}\mathbf{(r,}t)$ being the scalar kinetic pressure, i.e.,
\begin{equation}
p_{1}(\mathbf{r,}t)=\int d\mathbf{v}\frac{u^{2}}{3}f(\mathbf{x,}t),
\label{moment-3}
\end{equation}%
Requiring, $\nabla p\mathbf{(r,}t)=\nabla p_{1}\mathbf{(r,}t)$ and $p_{1}%
\mathbf{(r,}t)$ strictly positive, it follows that $P_{o}$ is an arbitrary
strictly positive function of time, to be defined so that the physical
realizability condition $p\mathbf{(r,}t)\geq 0$ is satisfied everywhere in $%
\overline{\Omega }\times I$ ($I\subseteq \mathbb{R}$ being generally a
finite time interval);

\item $\left\{ \rho _{o},\mathbf{V,}p\right\} $ are advanced in time by
means of the inverse kinetic equation Eq.(\ref{inverse kinetic equation});

\item By appropriate choice of the mean-field force $\mathbf{F}$, the moment
equations can be proven to satisfy identically INSE, and in particular the
Poisson equation for the fluid pressure, as well the appropriate initial and
boundary conditions (see Ref.\cite{Ellero2005});

\item The mean-field force\ $\mathbf{F}$ results, by construction, function
only of the velocity moments (\ref{third-moment}), i.e., $\left\{ \rho _{o},%
\mathbf{V,}p_{1},\mathbf{Q,}\underline{\underline{\mathbf{\Pi }}}\right\} ,$
to be denoted as \emph{extended fluid fields}$.$Here $\mathbf{Q}$ and\textbf{%
\ }$\underline{\underline{\mathbf{\Pi }}}$ are respectively the relative
kinetic energy flux (defined in the reference frame locally at rest with
respect to the fluid) and the pressure tensor%
\begin{equation}
\mathbf{Q}=\int d^{3}v\mathbf{u}\frac{u^{2}}{3}f,  \label{Q}
\end{equation}%
\begin{equation}
\underline{\underline{\mathbf{\Pi }}}=\int d^{3}v\mathbf{uu}f;  \label{PP}
\end{equation}

\item The Maxwellian kinetic distribution function $f_{M},$ defined by the
equation (\ref{local Maxwellian distribution}), results a particular
solution of the inverse kinetic equation (\ref{inverse kinetic equation}) if
and only if $\left\{ \rho ,\mathbf{V,}p\right\} $ satisfy INSE.
\end{itemize}

Let us now prove that the inverse kinetic equation defined above (\ref%
{inverse kinetic equation}) is non-unique, even in the particular case of
local Maxwellian kinetic equilibria, due to the non-uniqueness in the
definition of the mean-field force $\mathbf{F}\ $and the streaming operator $%
L(\mathbf{F})$. In fact, let us introduce the parameter-dependent vector
field $\mathbf{F}(\alpha )$
\begin{equation}
\mathbf{F}(\alpha )=\mathbf{F}+\alpha \mathbf{u}\cdot \nabla \mathbf{V-}%
\alpha \nabla \mathbf{V\cdot u}\equiv \mathbf{F}_{0}(\alpha )+\mathbf{F}_{1}
\end{equation}%
where $\mathbf{F\equiv F}(\alpha =0),$ $\alpha \in \mathbb{R} $ is arbitrary
and we have denoted

\begin{eqnarray}
&\text{ \ \ \ \ \ \ \ \ \ \ }&\left. \mathbf{F}_{0}(\alpha )=\mathbf{F}%
_{0}-\alpha \Delta \mathbf{F}_{0}\equiv \mathbf{F}_{0a}+\Delta _{1}\mathbf{F}%
_{0}(\alpha )\right. , \\
&&\Delta \mathbf{F}_{0}\equiv \mathbf{u}\cdot \nabla \mathbf{V-}\nabla
\mathbf{V\cdot u}, \\
&&\left. \Delta _{1}\mathbf{F}_{0}(\alpha )\equiv (1+\alpha )\mathbf{u}\cdot
\nabla \mathbf{V-\alpha }\nabla \mathbf{V\cdot u,}\right.
\label{eq.Delta_1_F_0}
\end{eqnarray}%
where $\mathbf{F}_{0}$ and $\mathbf{F}_{1}$ given by Eqs.(\ref{F0
non-maxwellian case}),(\ref{F1 non-Maxwellian case}). Furthermore, here we
have introduced also the quantity $\Delta _{1}\mathbf{F}_{0}(\alpha )$ to
denote the parameter-dependent part of $\mathbf{F}_{0}(\alpha )$. In fact,
it is immediate to prove the following elementary results:

a) for arbitrary $\alpha \in \mathbb{R},$ the local Maxwellian distribution (%
\ref{local Maxwellian distribution}) $f_{M}$ is a particular solution of the
inverse kinetic equation (\ref{inverse kinetic equation}) if and only if the
incompressible N-S equations are satisfied;

b) for arbitrary $\alpha $ in $\mathbb{R}$, the moment equations stemming
from the kinetic equation (\ref{inverse kinetic equation}) coincide with the
incompressible N-S equations;

c) the parameter $\alpha $ results manifestly functionally independent of
the kinetic distribution function $f(\mathbf{x},t).$

The obvious consequence is that the functional form of the vector field $%
\mathbf{F}_{0},$ and consequently $\mathbf{F,}$ which characterizes the
inverse kinetic equation (\ref{inverse kinetic equation}) is not unique. The
non-uniqueness in the contribution $\mathbf{F}_{0}(\alpha )$ is carried by
the term $\alpha \Delta \mathbf{F}_{0}$ which does not vanish even if the
fluid fields are required to satisfy identically INSE in the set $\Omega
\times I.$

We intend to show in the sequel that the value of the parameter $\alpha $
can actually be uniquely defined by a suitable prescription on the streaming
operator (\ref{streaming operator}) and the related mean-field force.

\section{A unique representation}

To resolve the non-uniqueness feature of the functional form of the
streaming operator (\ref{streaming operator}), due to this parameter
dependence, let us now consider again the inverse kinetic equation (\ref%
{inverse kinetic equation}). We intend to prove that the mean-field force $%
\mathbf{F,}$ and in particular the vector field $\mathbf{F}_{0}(\alpha )$,
can be given an unique representation in terms of a suitable set of fluid
fields $\left\{ \rho _{o},\mathbf{V,}p_{1},\mathbf{Q,}\underline{\underline{%
\mathbf{\Pi }}}\right\} $ defined by Eqs. (\ref{moment-1})-(\ref{moment-3})
and (\ref{Q}),(\ref{PP}), by introducing a symmetrization condition on the
mean field force $\mathbf{F}_{0}(\alpha ).$To reach this conclusion it is
actually sufficient to impose that the kinetic energy flux equation results
parameter-independent and suitably defined. Thus, let us consider the moment
equation which corresponds the kinetic energy flux density $G(\mathbf{x},t)=%
\mathbf{v}\frac{u^{2}}{3}$. Requiring that $f(\mathbf{x,}t)$ is an arbitrary
particular solution of the inverse kinetic equation (not necessarily
Maxwellian) for which the corresponding moment $\mathbf{q}=\int d^{3}v%
\mathbf{v}\frac{u^{2}}{3}f$ (kinetic energy flux vector) does not vanish
identically, the related moment equation takes the form%
\begin{eqnarray}
&&\text{ \ \ \ \ \ \ \ \ \ \ \ \ \ }\left. \frac{\partial }{\partial t}\int d%
\mathbf{v}G(\mathbf{x,}t)f+\nabla \cdot \int d\mathbf{vv}G(\mathbf{x,}%
t)f-\right.  \notag \\
&&\text{ \ \ \ \ \ \ \ \ \ \ \ \ }\left. -\int d\mathbf{v}\left[ \mathbf{F}%
_{0a}+\Delta _{1}\mathbf{F}_{0}(\alpha )+\mathbf{F}_{1}\right] \cdot \frac{%
\partial G(\mathbf{x,}t)}{\partial \mathbf{v}}f-\right. \\
&&\text{ \ \ \ \ \ \ \ \ \ \ \ \ }\left. -\int d\mathbf{v}f\left[ \frac{%
\partial }{\partial t}G(\mathbf{x,}t)+\mathbf{v\cdot \nabla }G(\mathbf{x,}t)%
\right] =0.\right.  \notag
\end{eqnarray}%
Introducing the velocity moments $p_{2}=\int d\mathbf{v}\frac{u^{4}}{3}f,$ $%
\underline{\underline{\mathbf{P}}}=\int d\mathbf{vuu}\frac{u^{2}}{3}f$ and $%
\underline{\underline{\underline{\mathbf{T}}}}=\int d\mathbf{vuuu}f,$ the
kinetic energy flux equation becomes therefore

\begin{eqnarray}
&&\text{ \ \ \ \ \ \ \ \ }\left. \frac{\partial }{\partial t}\mathbf{%
q+\nabla \cdot \underline{\underline{\mathbf{P}}}+\mathbf{\mathbf{V\cdot }}}%
\left[ \mathbf{\underline{\underline{\mathbf{P}}}+2\mathbf{VQ+VV}}p_{1}%
\right] \mathbf{+}\right.  \notag \\
&&\text{ \ \ \ \ \ \ \ }\left. +\left[ \frac{1}{\rho _{o}}\mathbf{f-}\nu
\nabla ^{2}\mathbf{V}\right] \cdot \left[ \mathbf{\underline{\underline{1}}}%
p_{1}+\frac{2}{3}\underline{\underline{\mathbf{\Pi }}}\right] -\right.
\notag \\
&&\text{ \ \ \ \ \ \ }\left. -\left[ \frac{3}{2}\mathbf{Q+V}p_{1}\right]
\left\{ \frac{D}{Dt}\ln p_{1}\mathbf{+}\frac{1}{p_{1}}\mathbf{\nabla \cdot Q-%
}\frac{1}{p_{1}^{2}}\left[ \mathbf{\nabla \cdot }\underline{\underline{%
\mathbf{\Pi }}}\right] \mathbf{\cdot Q}\right\} -\right.  \notag \\
&&\text{ \ \ \ \ \ }\left. -\frac{v_{th}^{2}}{2p_{1}}\mathbf{\nabla \cdot }%
\underline{\underline{\mathbf{\Pi }}}\cdot \left\{ \mathbf{\underline{%
\underline{1}}}\left[ \frac{p_{2}}{v_{th}^{2}}-\frac{3}{2}p_{1}\right] +%
\left[ \mathbf{QV+}\underline{\underline{\mathbf{P}}}\right] \frac{1}{%
v_{th}^{2}}-\frac{3}{2}\underline{\underline{\mathbf{\Pi }}}\right\} -\right.
\\
&\text{ \ \ \ \ \ \ }&\text{ \ \ \ \ \ }\left. -(1+\alpha )\mathbf{Q\cdot }%
\nabla \mathbf{V+\alpha \nabla \mathbf{V\cdot Q}}+\mathbf{\frac{2}{3}\nabla
\mathbf{V}:}\left( \mathbf{\underline{\underline{\underline{\mathbf{T}}}}+%
\underline{\underline{\mathbf{\Pi }}}V}\right) +\right.  \notag \\
&&\text{ \ \ \ \ \ \ }\left. +\frac{2}{3}\frac{\partial \mathbf{V}}{\partial
t}\cdot \underline{\underline{\mathbf{\Pi }}}+\frac{2}{3}\mathbf{V\cdot
\nabla V\cdot \underline{\underline{\mathbf{\Pi }}}+}\frac{2}{3}\mathbf{%
\nabla V:\underline{\underline{\underline{\mathbf{T}}}}=0.}\right.  \notag
\end{eqnarray}

Unlike the lower-order moment equations (obtained for $G(\mathbf{x,}t)=1,%
\mathbf{v,}u^{2}/3)$, the kinetic energy flux equation contains
contributions which depend linearly on the undetermined parameter $\alpha .$
These terms, proportional to the velocity gradient $\nabla \mathbf{V,}$
yield generally non-vanishing contributions to the rate-of-change of $%
\mathbf{q.}$ The sum of all such terms, which are carried respectively by $%
\Delta _{1}\mathbf{F}_{0}(\alpha ),$ $\mathbf{v\cdot \nabla }G(\mathbf{x,}t)$
and the convective term $\mathbf{\mathbf{V}}\cdot \nabla \mathbf{\mathbf{Q}}$%
, reads
\begin{equation}
\text{ \ \ \ }\mathbf{M}_{\alpha }(f)\equiv 2\mathbf{\nabla \cdot }\left(
\mathbf{V\mathbf{Q}}\right) -(1+\alpha )\mathbf{Q\cdot }\nabla \mathbf{V}+%
\mathbf{\alpha \nabla \mathbf{V\cdot Q}}.  \label{moment M_alfa}
\end{equation}%
which is the contribution to the rate-of-change of $\mathbf{q}$ which
results proportional both to $\mathbf{Q}$ (the relative kinetic energy flux)
and the velocity gradient (either $\nabla \mathbf{V}$ or $\mathbf{\mathbf{%
V\cdot }\nabla )}.$ In order to eliminate the indeterminacy of $\alpha $,
since $\alpha $ cannot depend on the kinetic distribution function $f,$ a
possible choice is provided by the assumption that $\mathbf{M}_{\alpha }(f)$
takes the symmetrized form
\begin{equation}
\mathbf{M}_{\alpha }(f)=2\mathbf{\nabla \cdot }\left( \mathbf{V\mathbf{Q}}%
\right) \mathbf{\mathbf{-}}\frac{1}{2}\mathbf{\nabla \mathbf{V\cdot Q+}}%
\frac{1}{2}\mathbf{Q\cdot \nabla V,}  \label{symmetruzation condition}
\end{equation}%
which manifestly implies $\alpha =1/2.$ Notice that the symmetrization
condition can also be viewed as a constitutive equation for the
rate-of-change of the kinetic energy flux vector. In this sense, it is
analogous to similar symmetrized constitutive equations adopted in customary
approaches to extended thermodynamics \cite{muller1998}. \ On the other
hand, Eq.(\ref{symmetruzation condition}) implies $\mathbf{M}_{\alpha }(f)=%
\frac{1}{2}\mathbf{\mathbf{Q\times }\xi ,}$ $\mathbf{\xi }=\nabla \times
\mathbf{V}$ being the vorticity field. Thus, $\mathbf{M}_{\alpha }(f)$ can
also be interpreted as the rate-of-change of the kinetic energy flux vector $%
\mathbf{Q}$ produced by vorticity field $\mathbf{\xi }$. From Eq.(\ref%
{symmetruzation condition}) it follows that $\mathbf{F}_{0}(\alpha )$ reads
\begin{eqnarray}
&&\left. \mathbf{F}_{0}(\alpha =\frac{1}{2})=-\frac{1}{\rho _{o}}\mathbf{f}%
+\right.   \label{new F0 non-maxwellian case} \\
&&\left. +\frac{1}{2}\left( \mathbf{u}\cdot \nabla \mathbf{V+}\nabla \mathbf{%
V\cdot u}\right) +\nu \nabla ^{2}\mathbf{V}.\right.
\end{eqnarray}%
As a consequence, the functional form of the streaming operator
results uniquely determined. Finally, for completeness, we notice
that the same representation for $\mathbf{F}_{0}(\alpha )$ can
also be obtained adopting the viewpoint described in Appendix B.
In fact, since $\alpha $ is functionally independent of the
kinetic distribution function $f(\mathbf{x},t),$ it can also be
defined in such a way to satisfy a suitable symmetry condition in
velocity-space, which \textit{holds in the particular case}
$f(\mathbf{x},t)=f_{M}(\mathbf{x},t)$. This is can be realized by
requiring that the Vlasov streaming operator $L(\mathbf{F})$
coincides in such a case with a suitable Fokker-Planck operator
with velocity-independent Fokker-Planck coefficients (see Appendix
B).

Finally, it is interesting to point out \ that the Vlasov streaming operator
$L(\mathbf{F})$ defined in terms of $\mathbf{F}_{0}(\alpha =\frac{1}{2})$
results by construction Markovian. Hence, the position (\ref{new F0
non-maxwellian case}) does not conflict with the Pawula theorem \cite%
{Pawula1967,Pawula1967-2} which yields a sufficient condition for the
positivity of the kinetic distribution function $f.$ The condition of
positivity for the kinetic distribution function satisfying the inverse
kinetic equation (\ref{inverse kinetic equation}) which corresponds to the
definition (\ref{new F0 non-maxwellian case}) for $\mathbf{F}_{0}(\alpha )$
has been investigated elsewhere \cite{Ellero2006a}. In particular by
assuming that $f$ results initially strictly positive and suitably smooth,
one can prove that $f$ satisfies an H-theorem both for Maxwellian and
non-Maxwellian distributions functions.

As a result of the previous considerations, it is possible to establish the
following uniqueness theorem:

\textbf{THEOREM \ 1 -- Uniqueness of the Vlasov streaming operator }$L(%
\mathbf{F})$

\emph{Let us assume that:}

\emph{1) the fluid fields} $\left\{ \rho ,\mathbf{V,}p\right\} $ \emph{and
volume force density }$\mathbf{f(r,V},t)$ \emph{belong to the functional
setting (\ref{NFS}); }

\emph{2) the operator }$L(\mathbf{F}),$ \emph{defining the inverse kinetic
equation (\ref{inverse kinetic equation})}, \emph{has the form of the Vlasov
streaming operator (\ref{streaming operator});}

\emph{3) the solution, }$f(\mathbf{x},t),$ \emph{of the inverse kinetic
equation (\ref{inverse kinetic equation}) exists, results suitably smooth in
}$\Gamma \times I$ \emph{and} \emph{its velocity moments }$\left\{ \rho _{o},%
\mathbf{V,}p_{1},\mathbf{Q},\underline{\underline{\mathbf{\Pi }}}\right\} $
\emph{define the fluid fields }$\left\{ \rho _{o},\mathbf{V,}p\right\} $%
\emph{\ which are classical solutions of INSE, together with Dirichlet
boundary conditions and initial conditions}$.$\emph{\ In addition, the
inverse kinetic equation admits, as particular solution, the local
Maxwellian distribution (\ref{local Maxwellian distribution});}

\emph{4) the mean-field force }$\mathbf{F}(\alpha )$\ \emph{is a
function only of the extended fluid fields } $\left\{ \rho
_{o},\mathbf{V,}p_{1},\mathbf{Q},\underline{\underline{\mathbf{\Pi
}}}\right\} ,$ \emph{while the parameter }$\alpha $\emph{\ does
not depend functionally on} $f(\mathbf{x},t);$

\emph{5) the vector field }$\Delta _{1}F_{0}(\alpha )$\emph{\ satisfies the
the symmetry condition (\ref{symmetruzation condition})}$.$

\emph{Then it follows that the mean-field force }$\mathbf{F}$\emph{\ in the
inverse kinetic equation (\ref{inverse kinetic equation}) is uniquely
defined in terms of}%
\begin{equation}
\mathbf{F}=\mathbf{F}_{0}+\mathbf{F}_{1},  \label{F Maxwellian-1}
\end{equation}%
\emph{where the vector fields }$\mathbf{F}_{0}$\emph{\ and }$\mathbf{F}_{1}$%
\emph{\ are given by Eqs. (\ref{new F0 non-maxwellian case}) and (\ref{F1
non-Maxwellian case}).}

\emph{PROOF}

Let us consider first the case in which the distribution function $f(\mathbf{%
x},t)$ coincides with the local Maxwellian distribution $f_{M}$ (\ref{local
Maxwellian distribution}). In this case by definition the moments $\mathbf{Q}%
,\underline{\underline{\mathbf{\Pi }}}$ vanish identically while, by
construction the mean mean-field force\emph{\ }is given by $\mathbf{F}%
(\alpha )$ [see Eq.()], $\alpha \in
\mathbb{R}
$ being an arbitrary parameter.

Let us now assume that $f(\mathbf{x},t)$ is non-Maxwellian and that its
moment $\mathbf{M}_{\alpha }(f)$ defined by Eq.(\ref{moment M_alfa}) is
non-vanishing. In this case the uniqueness of $\mathbf{F}$ follows from
assumptions 4 and 5. In particular the parameter $\alpha $ is uniquely
determined by the symmetry condition (\ref{symmetruzation condition}) in the
moment $\mathbf{M}_{\alpha }(f)$. Since by assumption $\alpha $ is
independent of $f(\mathbf{x},t)$ the result applies to arbitrary
distribution functions (including the Maxwellian case).

Let us now introduce the vector field $\mathbf{F}^{\prime }\mathbf{=F+}$ $%
\Delta \mathbf{F,}$ where the vector field $\Delta \mathbf{F}$ is assumed to
depend functionally on $f(\mathbf{x},t)$ and defined in such a way that:

A) the kinetic equation $L(\mathbf{F}^{\prime })f(\mathbf{x},t)=0$ yields an
inverse kinetic theory for INSE, satisfying hypotheses 1-5 of the present
theorem, and in particular it produces the same moment equation of the
inverse kinetic equation (\ref{inverse kinetic equation}) for $G(\mathbf{x,}%
t)=1,\mathbf{v,}E\equiv \frac{1}{3}u^{2}$;

B) there results identically $\Delta \mathbf{F}(f_{M})\equiv 0,$ i.e., $%
\Delta \mathbf{F}$ vanishes identically in the case of a local Maxwellian
distribution $f_{M}.$

Let us prove that necessarily $\Delta \mathbf{F}(f)\equiv 0$ also for
arbitrary non-Maxwellian distributions $f$ which are solutions of the
inverse kinetic equation. First we notice that from A and B, due to
hypotheses 3 and 4, it follows that $\Delta \mathbf{F}$ must depend linearly
on $\mathbf{Q},\underline{\underline{\mathbf{\Pi }}}-p_{1}\underline{%
\underline{\mathbf{1}}}.$ On the other hand, again due to assumption A the
vector field $\Delta \mathbf{F}$ must give a vanishing contribution to the
moments the kinetic equation evaluated with respect to $G(\mathbf{x,}t)=1,%
\mathbf{v,}E\equiv \frac{1}{3}u^{2}.$ Hence, in order that also $\mathbf{F}%
^{\prime }$ depends only on the moments $\left\{ \rho _{o},\mathbf{V,}p_{1},%
\mathbf{Q},\underline{\underline{\mathbf{\Pi }}}\right\} $ (hypothesis 4)
necessarily it must result $\Delta \mathbf{F}(f)\equiv 0$ also for arbitrary
non-Maxwellian distributions $f.$

\section{Fulfillment of the energy equation}

As a further development, let us now impose the additional requirement that
the inverse kinetic theory yields explicitly also the energy equation (\ref%
{energy equation}).

We intend to show that the kinetic equation fulfilling such a condition can
be obtained by a unique modification of the mean-field force $\mathbf{%
F\equiv F}_{0}\mathbf{(x,}t)+\mathbf{F}_{1}\mathbf{(x,}t),$ in particular
introducing a suitable new definition of the vector field $\mathbf{F}_{1}%
\mathbf{(x,}t)$ [Eq.(\ref{F1 non-Maxwellian case}), Appendix A]$.$ The
appropriate new representation is found to be

\begin{equation*}
\mathbf{F}_{1}\mathbf{(x,}t;f)=\frac{1}{2}\mathbf{u}\left\{ \frac{\partial
\ln p_{1}}{\partial t}-\frac{1}{p_{1}}\mathbf{V\cdot }\left[ \frac{\partial
}{\partial t}\mathbf{V}+\mathbf{V\cdot \nabla V}\right. \right. +
\end{equation*}%
\begin{equation*}
\left. +\frac{1}{\rho _{o}}\mathbf{f}-\nu \nabla ^{2}\mathbf{V}\right] +
\end{equation*}%
\begin{equation}
\left. +\frac{1}{p_{1}}\mathbf{\nabla \cdot Q}-\frac{1}{p_{1}^{2}}\left[
\mathbf{\nabla \cdot }\underline{\underline{\mathbf{\Pi }}}\right] \mathbf{%
\cdot Q}\right\} +  \label{new F1 non-maxwellian}
\end{equation}%
\begin{equation*}
+\frac{v_{th}^{2}}{2p_{1}}\mathbf{\nabla \cdot }\underline{\underline{%
\mathbf{\Pi }}}\left\{ \frac{u^{2}}{v_{th}^{2}}-\frac{3}{2}\right\} .
\end{equation*}%
As a consequence, the following result holds:

\textbf{THEOREM \ 2 -- Inverse kinetic theory for extended INSE}

\emph{Let us require that:}

\emph{1) assumptions 1-3 of Thm.1 are valid;}

\emph{2) the mean-field }$\mathbf{F}$ $\ $\emph{is defined:}%
\begin{equation}
\mathbf{F}=\mathbf{F}_{0}+\mathbf{F}_{1},  \label{F Maxwellian}
\end{equation}%
\emph{with }$\mathbf{F}_{0}$\emph{\ and} $\mathbf{F}_{1}$ \emph{given by
Eqs. (\ref{new F0 non-maxwellian case}) and (\ref{new F1 non-maxwellian}).}

\emph{Then it follows that:}

\emph{\ A) }$\left\{ \rho ,\mathbf{V,}p\right\} $ \emph{are classical
solutions of extended INSE in }$\Omega \times I$ \emph{[equations (\ref%
{INSE-1})-(\ref{INSE-6b}) and (\ref{energy equation})] if and only if \ the
Maxwellian distribution function }$f_{M}$\emph{\ (\ref{local Maxwellian
distribution}) is a particular solution of the inverse kinetic equation (\ref%
{inverse kinetic equation});}

\emph{B) provided that the solution }$f(\mathbf{x,}t)$\emph{\ of the inverse
kinetic equation (\ref{inverse kinetic equation}) exists in }$\Gamma \times
I $ \emph{and results suitably summable in the velocity space }$U,$ \emph{so
that the moment equations} \emph{of (\ref{inverse kinetic equation})
corresponding to the weight-functions }$G(\mathbf{x,}t)=1,\mathbf{v,}E\equiv
\frac{1}{3}u^{2}$ \emph{exist}$,$\emph{\ they} \emph{coincide necessarily
with extended INSE.}

\emph{C) the two representations} \emph{(\ref{F1 non-Maxwellian case}) and}
\emph{(\ref{new F1 non-maxwellian}) for }$\mathbf{F}_{1}$ \emph{coincide
identically }

\emph{PROOF:}

A) The proof is straightforward. In fact, recalling Thm.1 in \cite%
{Ellero2005}, we notice that Eqs. (\ref{new F1 non-maxwellian}) and (\ref{F1
non-Maxwellian case}) manifestly coincide if and only if \ the energy
equation (\ref{energy equation}) is satisfied identically, i.e., if the
fluid fields are solutions of extended INSE.

B) The first two moment equations corresponding to $G(\mathbf{x,}t)=1,%
\mathbf{v}$ are manifestly independent of the form of $\mathbf{F}_{1},$ both
in the case of Maxwellian and non-Maxwellian distributions, i.e., (\ref{new
F1 non-maxwellian}) and (\ref{F1 non-Maxwellian case}). Hence, in such a
case Thm.3 of \cite{Ellero2005} applies, i.e., the moment equations yield
INSE. Let us consider, in particular, the third moment equation
corresponding to $G(\mathbf{x,}t)=\frac{1}{3}u^{2}$ ,%
\begin{equation}
\frac{\partial }{\partial t}p_{1}+\nabla \cdot \mathbf{Q}+\nabla \cdot \left[
\mathbf{V}p_{1}\right] -\frac{2}{3}\int d\mathbf{vF(x,}t)\mathbf{u}f+\frac{2%
}{3}\mathbf{\nabla V:\underline{\underline{\Pi }}}=0.  \label{third-moment}
\end{equation}%
Invoking Eqs. (\ref{new F0 non-maxwellian case}) and (\ref{new F1
non-maxwellian}) for $\mathbf{F}_{0}$ and $\mathbf{F}_{1},$ Eq.(\ref%
{streaming operator}) reduces to
\begin{equation*}
p_{1}\nabla \cdot \mathbf{V}=0
\end{equation*}%
if and only if the energy equation (\ref{energy equation}) is satisfied.
Since by construction $p_{1}>0,$ this yields the isochoricity condition (\ref%
{INSE-3}).

C) Finally, since\ thanks to A) $\left\{ \rho ,\mathbf{V,}p\right\} $ are
necessarily classical solutions of INSE, it follows that they fulfill
necessarily also the energy equation (\ref{energy equation}). Hence, (\ref%
{F1 non-Maxwellian case}) and (\ref{new F1 non-maxwellian}) coincide
identically in \emph{\ }$\Gamma \times I$.

We conclude that (\ref{new F0 non-maxwellian case}) and (\ref{new F1
non-maxwellian}) provide a new form of the inverse kinetic equation applying
also to non-Maxwellian equilibria, which results alternative to that given
earlier in \cite{Ellero2005}. The new form applies necessarily to classical
solutions. Since weak solutions (and hence possibly also numerical
solutions) of INSE may not satisfy exactly the energy equation, the present
inverse kinetic theory based on the new definition given above [see Eq.(\ref%
{F1 non-Maxwellian case})] for the vector field $\mathbf{F(x,}t)$ provides a
necessary condition for the existence of strong solutions of INSE. The
result seems potentially relevant both from the conceptual viewpoint in
mathematical research and for numerical applications.

\section{Conclusions}

In this paper the non-uniqueness of the definition of the inverse kinetic
equation defined by Ellero and Tessarotto (see \cite{Ellero2005}) has been
investigated, proving that the mean-field force $\mathbf{F}$ characterizing
such an equation depends on an arbitrary real parameter $\alpha .$ To
resolve the indeterminacy, a suitably symmetrization condition has been
introduced for the kinetic energy flux moment equation. As a consequence,
the functional form the mean-field force $\mathbf{F}$ which characterizes
the inverse kinetic equation results uniquely determined.

Finally, as an additional development, we have shown that, consistently with
the assumption that the fluid fields are strong solutions of INSE, the
mean-field force can be expressed in such a way to satisfy explicitly also
the energy equation.

The result appears significant from the mathematical viewpoint, the physical
interpretation of the theory and potential applications to the investigation
of complex fluids, such as for example those treated in \cite%
{ellero1,ellero2}). In fact, it proves that the inverse kinetic theory
developed in \cite{Ellero2005} can be given an unique form which applies to
classical solutions of INSE.

\smallskip

\textbf{ACKNOWLEDGEMENTS} The research was developed in the framework of the
PRIN Research Project "Modelli della teoria cinetica matematica nello studio
dei sistemi complessi nelle scienze applicate" (Italian Ministry of
University and Research). The authors are indebted with the reviewer for
useful comments.

\section{Appendix A: relevant velocity moments and mean-field force}

Here we recall for completeness the expressions of the velocity-moments of
the kinetic distribution function\ and of the mean-field force $\mathbf{F}$
given in \cite{Ellero2005}. The relevant moments are of the form
\begin{equation}
M(G)=\int d^{3}vG(\mathbf{x},t)f(\mathbf{x,}t),  \label{a-1}
\end{equation}
$G(\mathbf{x},t)$ being the weight functions $G(\mathbf{x},t)=1,\mathbf{v,}%
u^{2}/3,\mathbf{u}$ $u^{2}/3,$\textbf{\ }$\mathbf{u}$ $\mathbf{u.}$\textbf{\
}In particular, we identify the fluid fields $\left\{ \rho \equiv \rho _{o},%
\mathbf{V,}p\right\} $ with%
\begin{equation}
\rho _{o}=\int d^{3}vf(\mathbf{x,}t),  \label{a-5}
\end{equation}%
\begin{equation}
\mathbf{V}(\mathbf{r,}t)=\frac{1}{\rho }\int d^{3}v\mathbf{v}f(\mathbf{x,}t),
\label{a-6}
\end{equation}%
\begin{equation}
p\mathbf{(r,}t)=p_{1}\mathbf{(r,}t)-P_{o},  \label{a-7}
\end{equation}
$P_{o}$ being a strictly positive real constant defined so that the physical
realizability condition $p\mathbf{(r,}t)\geq 0$ is satisfied everywhere in
the closure of the fluid domain $\overline{\Omega }$. Finally the expression
of the mean-field force reads:
\begin{equation}
\mathbf{F(x,}t;f)=\mathbf{F}_{0}\mathbf{(x,}t;f)+\mathbf{F}_{1}\mathbf{(x,}%
t;f),  \label{F non-Maxwellian}
\end{equation}%
where $\mathbf{F}_{0}$ and $\mathbf{F}_{1}$ are the vector fields:

\begin{equation}
\mathbf{F}_{0}\mathbf{(x,}t;f)=\frac{1}{\rho _{o}}\left[ \mathbf{\nabla
\cdot }\underline{\underline{\mathbf{\Pi }}}-\mathbf{\nabla }p_{1}-\mathbf{f}%
\right] +\mathbf{u}\cdot \nabla \mathbf{V+}\nu \nabla ^{2}\mathbf{V,}
\label{F0 non-maxwellian case}
\end{equation}%
\begin{equation}
\mathbf{F}_{1}\mathbf{(x,}t;f)=\frac{1}{2}\mathbf{u}\left\{ \frac{D}{Dt}\ln
p_{1}\mathbf{+}\frac{1}{p_{1}}\mathbf{\nabla \cdot Q-}\frac{1}{p_{1}^{2}}%
\left[ \mathbf{\nabla \cdot }\underline{\underline{\mathbf{\Pi }}}\right]
\mathbf{\cdot Q}\right\} +  \label{F1 non-Maxwellian case}
\end{equation}%
\begin{equation*}
+\frac{v_{th}^{2}}{2p_{1}}\mathbf{\nabla \cdot }\underline{\underline{%
\mathbf{\Pi }}}\left\{ \frac{u^{2}}{v_{th}^{2}}-\frac{3}{2}\right\} ,
\end{equation*}%
where the moments $p_{1},\mathbf{Q}$ and $\underline{\underline{\mathbf{\Pi }%
}}$ are given by Eqs.(\ref{moment-1}) - (\ref{moment-3}). In particular, for
the Maxwellian kinetic equilibrium (\ref{local Maxwellian distribution})
there results identically
\begin{eqnarray}
&\text{ \ \ \ \ \ \ \ \ \ \ \ \ \ \ \ \ }&\left. \underline{\underline{%
\mathbf{\Pi }}}=p_{1}\underline{\underline{\mathbf{1}}},\right.  \label{a-8}
\\
&&\left. \text{ }\mathbf{Q}\mathbf{=0}.\right.  \label{a-9}
\end{eqnarray}%
\smallskip

\section{Appendix B: Fokker-Planck representation}

It is interesting to point out that the choice of the parameter $\alpha =1/2$
and of the mean-field force $\mathbf{F}_{0}(\alpha )$ (\ref{new F0
non-maxwellian case}) can be obtained also by requiring that the Vlasov
streaming operator (\ref{streaming operator}) results "equivalent" to an
appropriate Fokker-Planck operator (i.e., that it yields an inverse kinetic
theory for INSE which admits local Maxwellian equilibria).

Since by assumption the parameter $\alpha $ does not depend functionally on $%
f(\mathbf{x},t)$ it is sufficient to impose its validity \textit{only} in
the case of local Maxwellian equilibria $f_{M}.$ In such a case, it is
possible to require that there results identically
\begin{equation}
L(\mathbf{F})f_{M}=\left( \frac{\partial }{\partial t}+\mathbf{v\cdot }\frac{%
\partial }{\partial \mathbf{r}}\right) f_{M}+\sum\limits_{i=1,2,3}\frac{%
\partial ^{i}}{\left( \partial \mathbf{v}\right) ^{i}}\left\{ \mathbf{D}%
_{i}f_{M}\right\} \equiv L_{FP}(\mathbf{F})f_{M},
\label{equivalent representation}
\end{equation}%
where the Fokker-Planck coefficients $\mathbf{D}_{i}$ are assumed
velocity-independent and $L_{FP}(\mathbf{F})$ denotes a Fokker-Planck
operator which by construction is equivalent to the Vlasov operator $L(%
\mathbf{F}).$ There results
\begin{equation}
\mathbf{D}_{1}(\mathbf{\mathbf{r,}}t;f_{M})=-\frac{1}{\rho _{o}}\mathbf{f}%
+\nu \nabla ^{2}\mathbf{V},
\end{equation}%
\begin{equation}
\underline{\underline{\mathbf{D}}}_{2}(\mathbf{\mathbf{r,}}t;f_{M})=-\frac{%
p_{1}}{\rho _{o}}\left\{ \frac{1}{2}\nabla \mathbf{V+}\underline{\underline{%
\mathbf{I}}}\frac{1}{2}\left[ \frac{\partial \ln p_{1}}{\partial t}+\mathbf{%
V\cdot \nabla }\ln p_{1}\right] \right\} ,
\end{equation}%
\begin{equation}
\underline{\underline{\underline{\mathbf{D}}}}_{3}(\mathbf{\mathbf{r,}}%
t;f_{M})=\frac{1}{2}\underline{\underline{\mathbf{I}}}\left( \frac{p_{1}}{%
\rho _{o}}\right) ^{2}\nabla \ln p_{1}.
\end{equation}%
On the other hand, the relationship (\ref{equivalent representation}) holds
if and only if
\begin{equation}
\mathbf{F}(\alpha )=\mathbf{D}_{1}-\frac{\rho _{o}}{p_{1}}\left[ \mathbf{%
u\cdot }\underline{\underline{\mathbf{D}}}_{2}+\underline{\underline{\mathbf{%
D}}}_{2}\cdot \mathbf{u}\right] -\left[ 2\frac{\underline{\underline{\mathbf{%
1}}}}{v_{th}^{2}}+4\frac{\mathbf{uu}}{v_{th}^{2}}\right] :\underline{%
\underline{\underline{\mathbf{D}}}}_{3}(\mathbf{\mathbf{r,}}t),
\end{equation}%
namely the parameter $\alpha $ necessarily results equal to $1/2$, i.e.,
\begin{equation}
\mathbf{F}_{0}(\alpha =\frac{1}{2})=-\frac{1}{\rho _{o}}\mathbf{f}+\frac{1}{2%
}\left( \mathbf{u}\cdot \nabla \mathbf{V+}\nabla \mathbf{V\cdot u}\right)
+\nu \nabla ^{2}\mathbf{V.}
\end{equation}

\end{document}